\documentclass{IEEEtran}
\usepackage{amsmath,amssymb}
\usepackage{algorithmic}
\usepackage{graphicx}
\usepackage{textcomp}
\usepackage{subfig}
\usepackage{caption}
\usepackage{tikz}
\usepackage{pgfplotstable}
\usepackage[english]{babel}
\pgfplotsset{compat=1.14}
\usetikzlibrary{patterns}
\usepackage[adjust]{adjustbox}
\definecolor{hous}{HTML}{b88b4d}
\definecolor{green}{HTML}{79c561}
\definecolor{farming}{HTML}{ded94c}
\definecolor{trans}{HTML}{5089a0}
\definecolor{indu}{HTML}{76dded}
\definecolor{other}{HTML}{91726e}
\definecolor{aku}{HTML}{db7c79}
\definecolor{water}{HTML}{7982db}
\definecolor{techinfra}{HTML}{303355}
\usepackage[usestackEOL]{stackengine}
\usepackage[backend=bibtex,style=numeric,sorting=none]{biblatex}
\renewbibmacro{in:}{}
\def\BibTeX{{\rm B\kern-.05em{\sc i\kern-.025em b}\kern-.08em
    T\kern-.1667em\lower.7ex\hbox{E}\kern-.125emX}}
\begin{document}
\title{C-Balancer: A System for Container Profiling and Scheduling}
\author{\IEEEauthorblockN{Akshay Dhumal\IEEEauthorrefmark{1},
Dharanipragada Janakiram\IEEEauthorrefmark{2}}
\\
\IEEEauthorblockA{Department of Computer Science,
Indian Institute of Technology Madras\\
Email: \IEEEauthorrefmark{1}akshayd@cse.iitm.ac.in,
\IEEEauthorrefmark{2}djram@iitm.ac.in}}

\maketitle
\begin{abstract}
Linux containers have gained high popularity in recent times. This popularity is significantly due to various advantages of containers over Virtual Machines (VM). The containers are lightweight, occupy lesser storage, have fast boot up time, easy to deploy and have faster auto-scaling.  The key reason behind the popularity of containers is that they leverage the mechanism of micro-service style software development, where applications are designed as independently deployable services. There are various container orchestration tools for deploying and managing the containers in the cluster. The prominent among them are Docker Swarm, and Kubernetes. However, they do not address the effects of resource contention when multiple containers are deployed on a node. Moreover, they do not provide support for container migration in the event of an attack or increased resource contention. To address such issues, we propose C-Balancer, a scheduling framework for efficient placement of containers in the cluster environment. C-Balancer works by periodically profiling the containers and deciding the optimal container to node placement. Our proposed approach improves the performance of containers in terms of resource utilization and throughput. Experiments using a workload mix of Stress-NG and iPerf benchmark shows that our proposed approach achieves a maximum performance improvement of 58\% for the workload mix. Our approach also reduces the variance in resource utilization across the cluster by 60\% on average. 
\end{abstract}


\section{Introduction}
Virtualization is seen as a backbone for High Performance Computing (HPC). Virtualization involves abstracting the various resources of a system using a software layer. The hypervisor based virtualization creates a nearly identical system using the virtualized resources. These virtualized systems are called Virtual Machines(VM).
Several studies \cite{felter2015updated} , \cite{soltesz2007container}, \cite{ruiz2015performance}  have shown that VMs suffer from higher performance overhead, larger storage, slower scalability and are heavier to migrate. Container based virtualization also called as Operating System (OS) based virtualization, evolved as an alternative to hypervisor based virtualization, by abstracting the underlying OS as a resource. In contrast to a VM, the containers share a common OS running on the physical machine. The sharing of the Operating System facilitates the containers to be extremely light-weight, occupy lesser storage space, scalable and faster to deploy. The containers have emerged as a lucrative option for providing services in public as well as private clouds.

The resource isolation for a container is provided by the kernel namespaces \cite{biederman2006multiple}. These namespaces allow different processes to have different views of the system. The resources that can be grouped under namespaces are process PID, file system mount points, interprocess communication, network, etc. The resource management for a container is done by control groups (cgroups) \cite{cgroup}, that limits the resource usage per process group. Using cgroups, it is possible to bound the resources (CPU, memory, I/O) of a container. The OS based container virtualization significantly gained mass attention and adoption with the release of Docker, which is a container orchestration and management tool designed to simplify the creation and deployment of containers on a single host. With the growing usage and popularity, the containers were seen as a replacement for VM in public and private cloud operations. Several tools provided container orchestration in a multi-node setup, such as Kubernetes \cite{google}, Apache Mesos \cite{hindman2011mesos}, etc. Docker \textit{Swarm} \cite{swarm} emerged as one of the popular tools for deploying, managing and scheduling containers on multiple nodes in a cluster.

Swarm uses the master-slave architecture to orchestrate containers across multiple nodes. The containers are deployed in the Swarm cluster as a \textit{service}. A \textit{service} is defined as a task with an associated state that includes the number of replicas, exposed ports, configurations etc. The task runs as a separate container in a cluster. Effectively, running multiple tasks or replicas is equivalent to launching multiple containers in the Swarm cluster. The service definition to launch the containers is provided to the Swarm \textit{ Manager}, which deploys the containers in the cluster based on certain scheduling policy.
There are mainly three strategies used for scheduling the containers in the Swarm cluster.
\begin{enumerate}
	\setlength \itemsep{0.5em}
    \item \textit{Spread}: This approach aims to balance the utilization of the worker nodes by deploying an equal number of containers on all nodes. Effectively, the strategy places the container on the node with the least number of active running containers. 
    \item \textit{Bin pack}:
    In this approach, a container is placed on the most packed node. Evidently, the smallest node gets to deploy the container which satisfies the container's resource requirement. The strategy tries to pack as many containers on a node so as to minimize the number of active worker nodes.
    \item \textit{Random}: As the name suggests, this scheduling policy deploys the container on a randomly chosen worker node. 
\end{enumerate}

Each of the scheduling policy has certain limitations. The \textit{Spread} strategy looks attractive but does not consider the run-time behaviour of a container to decide the schedule. The contention caused by shared resource utilization impacts the throughput of the container. For instance, executing  multiple  containers having a higher memory footprint on the same node may cause significant degradation in performance. The \textit{Bin-pack} strategy places containers on the most packed node thereby reducing the number of nodes. As this strategy places the containers on the heavily loaded node, the containers suffer from shared resource contention, thereby hurting the application performance. 

\begin{figure}
\includegraphics[scale=0.25]{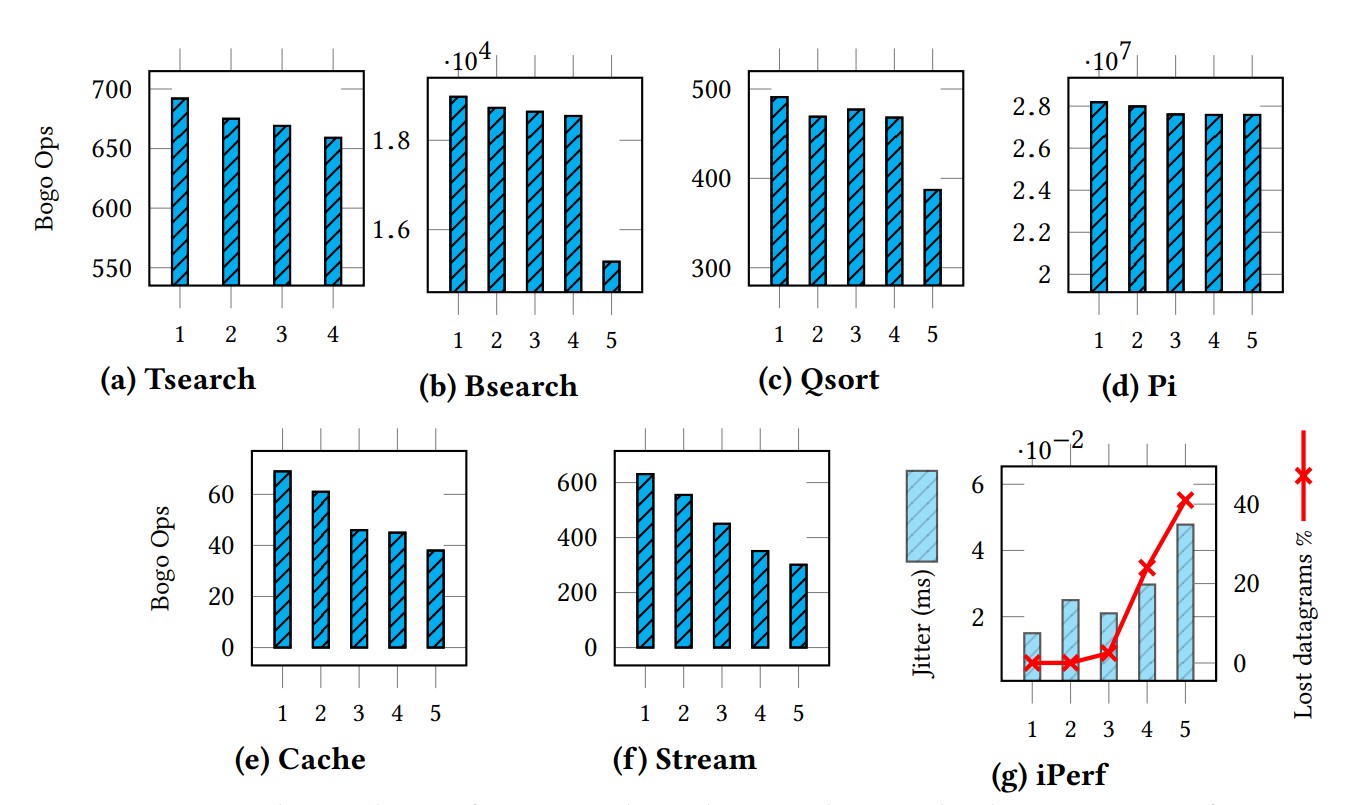}
\captionsetup{font=scriptsize,skip=1pt}	
\caption[]{The performance degradation when multiple containers of same application are launched on a single node for Stress-NG and iPerf benchmark. The x axis denotes the number of containers and y axis shows the throughput in Bogo Ops. }
\label{fig:contention}
\vspace{-6mm}
\end{figure}
Moreover, the container placement strategies do not migrate the containers in the event of an attack or increased resource contention. We argue why migrating a container is essential,
\begin{enumerate}
\setlength \itemsep{0.5em}
\item The default container placement strategy does not intelligently decide the schedule, that can have a significant impact on the performance of the application. For instance, placing containers incurring heavy cache or memory bandwidth on the same worker node, can degrade the performance. Likewise, scheduling containers having high I/O usage on the same worker node increases the latency of disk accesses, thereby impacting application performance. Figure \ref{fig:contention} shows the performance degradation when multiple  containers of the same application are scheduled on the same node. We evaluated the impact of the resource contention on the performance of the application. We containerize the Stress-NG \cite{Stress-NG} and iPerf \cite{iperf3} benchmark programs for the experimentation. We execute each of the benchmark programs for a total of 60 seconds. The iPerf client stresses the iPerf server with a maximum of 400 Mbits per second bandwidth. The performance impact for CPU bound jobs (\texttt{pi}) was significantly less than that of cache and memory bound programs (\texttt{Tsearch, Cache, Stream}). We also noticed a significant network performance degradation with the iPerf benchmark. There is an overall increase in the jitter values and lost datagrams, with the increase in the number of iPerf client containers.  
\item Docker Swarm employs the \textit{Spread} strategy to schedule containers in the cluster. The strategy chooses a worker node randomly, whenever there are an equal number of active containers across the worker nodes. In such scenarios, the \textit{Spread} strategy simply transforms to the \textit{Random} strategy resulting in cluster instability. Therefore, it becomes critical to re-balance and stabilize the cluster. Moreover, the order of container launch becomes critical in such scenarios. 
\item If the containers are killed or stopped accidentally, the \textit{Manager} relaunches them based on its container \textit{restart policy}. The erratic behaviour of the \textit{Manager} results in a significant loss of data and computation. The re-launch can have notable performance impact for containers running applications where pre-processing is required. Consider a web search application that typically involves indexing the database before starting the web-search service. The indexing (pre-processing) step is a massive operation and requires extensive computation. Restarting the containers in a cluster without preserving the computation, can lead to significant performance penalty, lesser availability and can even lead to the violation of the Service Level Agreements (SLAs). Such use cases require freezing the container state and restarting it on a different node. 
\end{enumerate}

The container orchestration tools should, therefore, encapsulate the container migration as a preliminary service. The primary concerns for implementing such a service should address various factors such as the effects of container migration, container migration time and the frequency of container migrations. There are several such aspects which require careful analysis. First, we need to define a specific scheduling objective for containers in the cluster. The objectives may vary from increased system stability, higher performance, reduced resource contention to stringent adherence of the SLAs. Second, the container needs to be migrated to the correctly chosen target node to meet the scheduling objective. Third, high availability must be ensured by minimizing the total downtime of a container. The downtime caused due to migration can severely increase the user experienced latency. Therefore the service hand-off must be seamless and should provide on-the-fly container provisioning. Fourth, the migration should aim to reduce the total data transfer incurred. 
Compromising on any of the factors may severely impact the performance of the cluster. A wrong container placement decision or a high migration time can eventually be catastrophic for applications running on the containers. 

In this paper, we propose a novel framework \textit{C-Balancer}, which leverages the runtime metrics of the containers to find an optimal container placement. We also use the layered file system of Docker image to speed up the process of container migration.

The rest of the paper is organized as follows:
Section \ref{sect:migration} describes the various approaches to container migration in detail. Section \ref{sect:framwork} describes the architecture of \textit{C-Balancer} and empirically evaluates the approach. Section \ref{sect:eval} presents the evaluation. Section \ref{sect:related} explores the related work. Finally we conclude the paper along with future work in section \ref{sect:concl}.

\section{Container Migration}
\label{sect:migration}
Docker is a container orchestration tool and has gained mass popularity in recent years. It enables us to package and run an application in an isolated environment known as a container. Containers are the runtime entities and are created using the Docker image that consists of multiple layers. To build a Docker image, we need to create a \texttt{Dockerfile} and specify the commands to create an image. The creation of an image starts with a base layer. With every new command, a layer is added on top of the existing layers. Essentially, a layer is just some files generated by executing the image creation commands. Each layer has a unique SHA256 identification. The layers are stacked upon one another to create a Docker image. The layers of Docker image are read-only and therefore cannot be altered. When a container is launched using an image, a new thin writable layer called \texttt{init} is created on top of the existing base layers. The \texttt{init} layer does not persist permanently on the disk and gets removed whenever the container is deleted. The \text{init} layer is required because any modifications to the file system by the  container processes must be committed to the disk. Consider a web server application, running as a container and writing logs to the file system. If the layers are read-only, it will inhibit the application to write logs to the disk. However, with a new writable layer on top, the logs can be committed to the temporary \texttt{init} layer.
The container scheduling decisions are coupled with the task of container migration which involves freezing the running application and checkpointing it as a set of files on the persistent storage. 
Checkpointing is a process that involves saving all essential runtime attributes of a process, so that the process can be restarted again. It includes saving the process tree, threads, open files, network sockets, memory maps, file descriptors, pages, intermediate values in registers, mount-points, etc. Checkpoint/Restore in Userspace (CRIU) \cite{criu} is a tool developed for Linux Operating system checkpointing and restoring the normal processes. However, there are certain limitations of CRIU, which restricts its seamless usage for the Docker containers. CRIU does not checkpoint the file system of a container. Consider a container performing write operations on the file system. These write operations are performed on the top \texttt{init} layer, causing changes to the file pointers.
The file pointers are saved by CRIU during checkpoint operation. However, the files that are modified are not persisted to the disk. The modified files are part of a temporary \texttt{init} layer of the container image. As a result, during the process restoration, the file pointers are recovered, whereas the files are not. This results in process restoration error and  the container fails to restart.  So in order to fully restore a container, we need to checkpoint the container file system along with its process tree and memory. In the next subsection, we discuss the steps needed to perform the container migration.

\subsection{Steps for migrating a Container}
\begin{enumerate}
\setlength \itemsep{0.4em}
\item Once the migration request is initiated, the instructions are sent to the host node specifying the container identification and the address of the target node.
\item The next step involves check-pointing the container (process tree and memory) as a collection of files on the persistent storage.  The container is stopped, once the checkpointing operation is completed. We use \texttt{docker checkpoint create} command to checkpoint the container. Docker internally uses \texttt{CRIU dump} to checkpoint the processes. CRIU dumps the process tree of a container along with the memory pages, page-map information, file descriptor information, shared memory, etc.
This step also involves fetching the container metadata, that includes the information about the mount points, exposed ports, entry-point instructions and the allocated resources such as CPU quota, memory limits, I/O bandwidth, etc. The container metadata is also transferred to the target node.  
\item As the container is check-pointed, the next step involves transferring the checkpoint data to the target node. The checkpoint data may require a larger storage space (in order of few mega bytes). This is so because the checkpoint data includes the memory pages owned by the container processes. The higher the memory footprint of the container, the higher the size  of checkpoint data of the container.
Transferring the checkpoint data across network incurs a high bandwidth and increased transfer time. To reduce overall checkpoint size, we compress the checkpoint data as a tar file. Compressing the checkpoint data, significantly reduces the checkpoint data size resulting in faster transfer time and reduced network bandwidth utilization. We evaluate the various factors affecting the container checkpoint time in section \ref{sect:factorschk}.
\item As the container is stopped at this moment, the file system of container would not change. The next step involves synchronizing the associated data storage of the container with the target node.
% The aim to to save the modified file system of the container.  
We discuss two different ways to accomplish this step in section \ref{sect:ap1}. 
\item The next step is to transfer the modified file system of the container to the target host. Section \ref{sect:ap1} describes this step in detail.
\item Once the checkpoints and file system of the container is available at the target node, we replicate the host container at the target node using the container metadata. 
\item As the container is created, we initiate the container by restoring the process tree on the target node using the container checkpoint data. We use \texttt{docker start} with \texttt{checkpoint} attribute to restore the container. Docker internally uses the \texttt{CRIU restore} command to restart the container. This step marks the end of container migration.
\end{enumerate}
We propose that the container file system synchronization between the host and the target node can be accomplished using two different methods.
\subsection{Approach 1: Freezing the entire file system of container}
\label{sect:ap1}

In this approach, we transfer the entire file system of the container as an archive file using the \texttt{Docker export} command. As discussed, the Docker uses a layered file system for the Docker images. These layers being read-only are not modified when the containers are launched using images. Any modifications to the file systems are made to the top \texttt{init} layer. Therefore, exporting the file system of the container includes exporting all the layers, including the \texttt{init} layer. On the target node, we use \texttt{Docker import} command to create a container image from the archive file.  The size of the entire archive file equals to the sum of the size of the Docker image and the \texttt{init} layer. The \texttt{init} layer generally occupies a lesser storage space and therefore the size of the container file system is equivalent to that of its image. Also, the use of \texttt{Docker export/import} command does not transfer the contents of the volumes mounted on a container.
\subsection{Approach 2: Freezing only the modified file system of container.}
\label{sect:ap2}
\begin{figure}
	\includegraphics[scale=0.25]{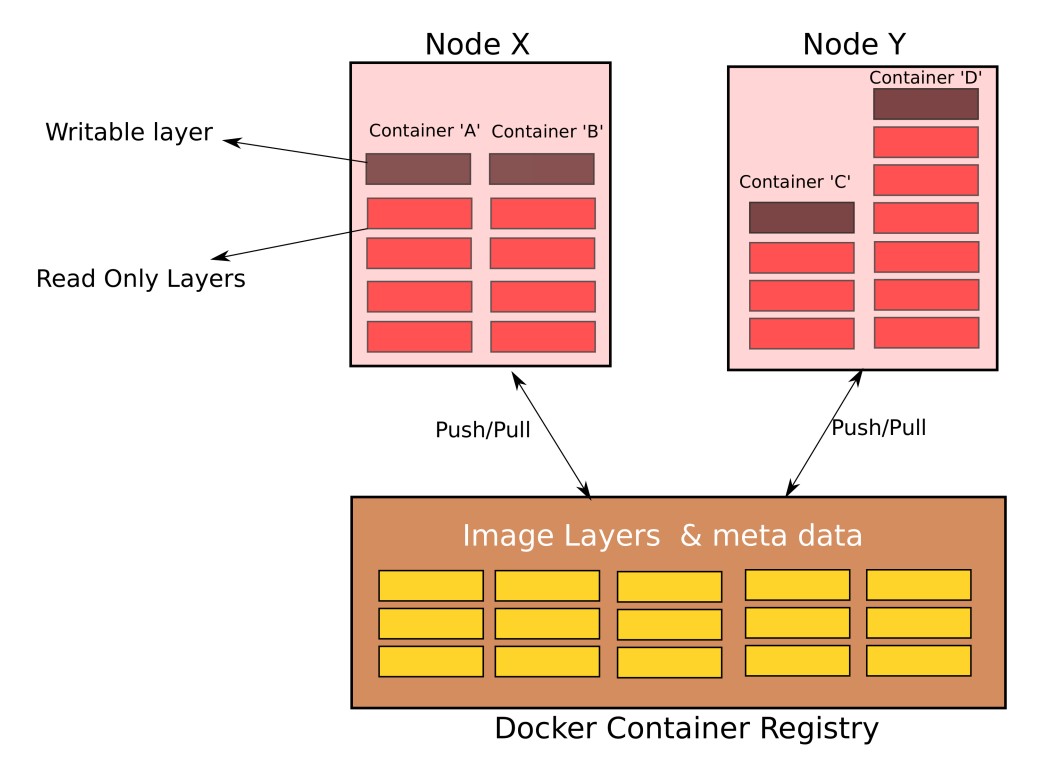}
	\vspace*{2mm} \captionsetup{font=scriptsize,skip=1pt} \caption[]{Approach 2}
	\vspace{-5mm}
	\label{fig:registry}
\end{figure}

In this approach, we only transfer the modified file system of a container to the target node.  The modifications to the file system are solely committed to the newly created  \texttt{init} layer of the container. In this approach, we create a  private registry to store  the container images. A registry is a collection of files and directories of each layer of the Docker image. The layers are stored as \texttt{tar} files in the registry. Each Docker image is associated with its manifest file containing information pertaining to its layers. It includes the hash (SHA256) of the layer, the size of the layer and the information of its parent layer.

We leverage the advantages of the layered file system of Docker containers. We push the layers of a Docker container, including \texttt{init} to the private registry using the \texttt{Docker push} command. During the push operation, the Docker daemon checks the \texttt{manifest} file to verify the presence of the image layers in the registry. If the layers are missing, the \texttt{push} command transfers the layers and updates the manifest file. However, if the image layers are already present, no actual data transfer occurs. At the target node, we use \texttt{pull} command to get the image. The Docker daemon at the target node also checks the manifest file for missing image layers. If the  layers are not present at the target node, the \texttt{pull} command fetches them from the registry and updates the manifest file. For a container image, only the top \texttt{init} layer is modified, keeping the remaining layers intact. The use registry thus enables us to transfer fewer layer to the target node. The approach considerably reduces the data transfer and speeds up the process of storage synchronization of a container.

Consider a scenario where the layers of an image are missing in the private registry. When the container is migrated for the first time, all the image layers gets transferred to the registry. However the next time a container relocates, only the top \texttt{init} layer is transferred to the registry as the remaining image are already present. \citeauthor{nathan2017comicon} \cite{nathan2017comicon} shows that nearly 50\% of the total image layers are shared across many popular Docker images. Therefore, with time as the registry gets filled up with the image layers, fewer layers get transferred every time a container migrates.  We use \texttt{commit} command to create a new Docker image of a running container. Figure \ref{fig:registry} depicts the use of the registry in our framework. We compare the two approaches for container file system synchronization in section \ref{sect:eval}.

\section{C-Balancer Framework}
\label{sect:framwork}
\begin{figure} 
	\centering
	\includegraphics[scale=0.3]{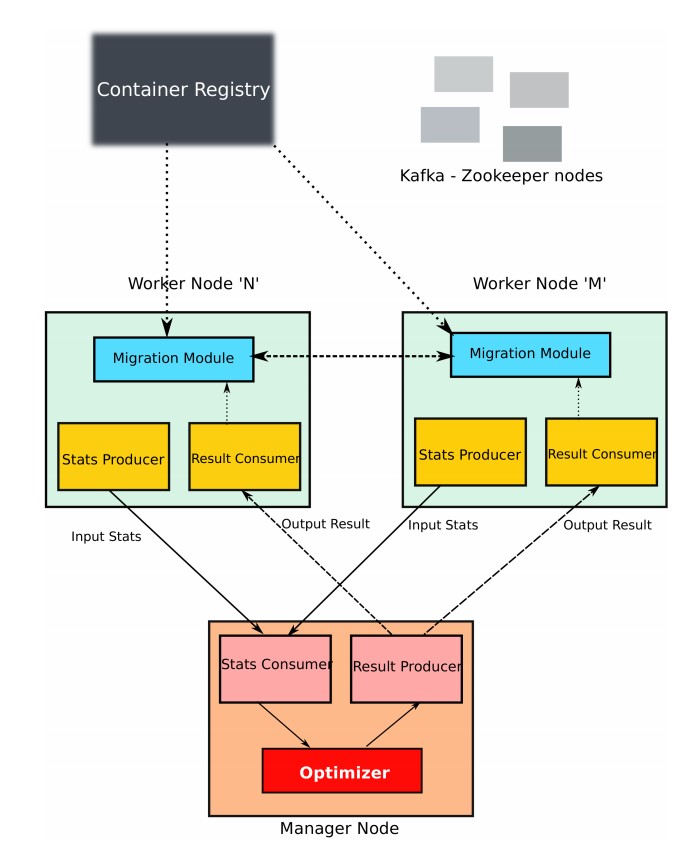}
	\captionsetup{font=scriptsize,skip=1pt}	
	\vspace{1mm}\caption[]{C-Balancer Framework}
	\label{fig:Framework}
\end{figure}
In this section, we present our proposed C-Balancer framework, which schedules and rebalances the containers across multiple nodes in a cluster. The idea is to capture and use run-time parameters of the container for making a scheduling decision. The run-time parameters of a container can be fetched from their respective cgroup. The number of run-time parameters exposed to a process using Hardware Performance Counters(HPC) are much more than those of a container. However, there exists a substantial number of parameters for a container that can be leveraged for making a container placement decision. There are various subsystems which capture and control the system resources of a cgroup. The runtime parameters obtained by these subsystems can be grouped as: 
\begin{itemize}
	\setlength \itemsep{0.3em}
	\item CPU Accounting (cpuacct): It reports the CPU resources consumed by the tasks in a cgroup such as \textit{usage},  \textit{stats}, etc.
	\item CPU sets (cpusets): It specifies the allocated CPUs and memory nodes for a cgroup such as \textit{cpus}, \textit{mems},  \textit{memory\_exclusive}, \textit{cpu\_exclusive} etc.
	\item Memory (memory): It reports the memory resources used by the tasks in a cgroup such as \textit{max\_usage\_in\_bytes}, \textit{stat},  \textit{numa\_stat}, etc.
	\item Block I/O (blkio): It reports the accesses to I/O devices by the tasks in a cgroup. It includes parameter such as \textit{sectors}, \textit{io\_service\_bytes}, etc.
	\item The network interface operates under the network namespace and hence the network parameters are fetched from the container namespace. 
\end{itemize}
We extend the idea of GAS \cite{janakiram2018gas} to a distributed cloud environment, where containers are the scheduling entities. Figure \ref{fig:Framework} presents the overall high-level architecture of our framework. Each node in the system is assigned a unique identification called \texttt{Node ID}. The framework has four major components: 
\begin{figure} 
	\centering
	\includegraphics[scale=0.25]{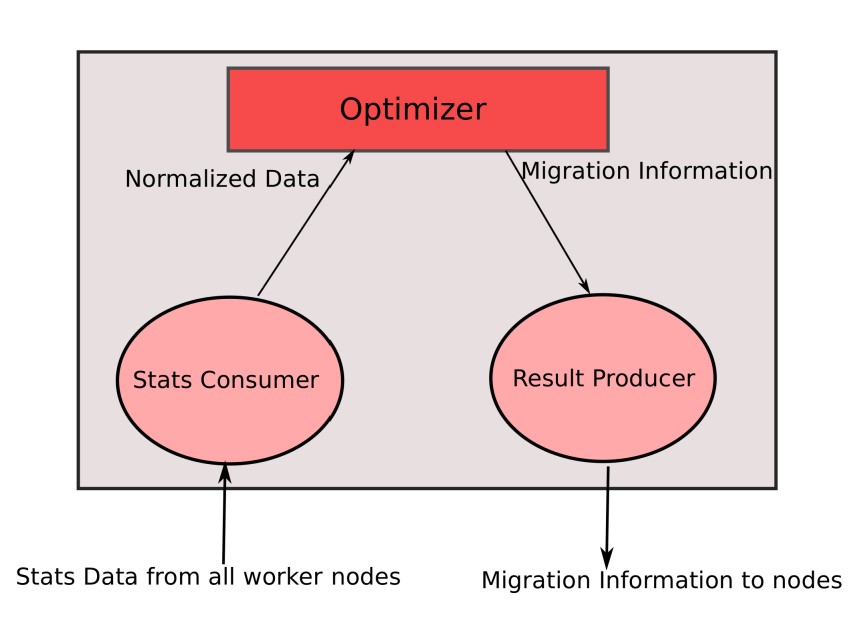}
	\vspace*{2mm} \captionsetup{font=scriptsize,skip=1pt}  
	\caption[]{Manager Node} \vspace{-5mm}
	\label{fig:manager}
\end{figure}

\begin{enumerate}
\setlength \itemsep{0.5em}

\item \textit{Kafka Broker - Zookeeper Nodes}: We use Apache Kafka \cite{kreps2011kafka}  publish-subscribe messaging system  to exchange data between various processes in our framework. Kafka is a distributed real-time streaming platform and provides capabilities to publish and subscribe to a stream of messages in a fault-tolerant, scalable and durable way. In our framework, Kafka provides interaction between multiple processes of  the \texttt{Manager} and \texttt{Worker} nodes wherein some processes generate the data and others consume it. The data generated by various processes are  stored in nodes called \textit{Kafka Broker}. The Kafka-Producer sends the messages, while the Kafka Consumers receives it.  The messages are published and subscribed from a category called \textit{topic}. The producers can, therefore, publish messages under a topic and consumers can subscribe to the topics to consume the messages. The worker node publishes messages under the topic \texttt{$M_x$}, where \texttt{$x$} stands for the worker node ID. The Manager node publishes message under the topic \texttt{$L_x$}, where \textit{$x$} stands for the worker node ID. The topic naming scheme enables the \texttt{Manager} to communicate messages with every \texttt{Worker} node. It also inhibits the Workers nodes to communicate with each other directly. For instance, if the manager needs to communicate a message to node \texttt{$x$}, it will publish the message under the topic \texttt{$L_x$}.  We used Kafka messaging system because the streams of messages are generated and consumed in real-time. Kafka internally uses Zookeeper \cite{zoo} to store and manage the Kafka meta-data. 

\item \textit{Manager Node}: It is responsible for coordinating all the activities for our \textit{C-Balancer} framework. Figure \ref{fig:manager} shows the various components of the \textit{Manager} node. The Manager periodically polls the \textit{Worker} node for container runtime metrics. The metrics are sent to the \textit{Optimizer}, which performs the optimization and decides the container to node placement. The Manager has primarily three different components: 
 \setlength{\parskip}{\baselineskip}
\begin{enumerate}
\setlength \itemsep{1em}
\item \textit{Stats Consumer}: It is responsible for collecting information and runtime container metrics of all the containers running across all the \textit{Worker} nodes in the cluster. It internally executes a Kafka Consumer, that consumes messages of \textit{Stats Producer} running across all the worker nodes. The data is processed and sent to the \textit{Optimizer}.

\item \textit{Result Producer}: The input to this component is a tuple \texttt{(Container Id, Host Node, Target Node)}. The host node is the node where the container is currently running and the target node is the destination node, where the container needs to be migrated. The job of \textit{Result Producer} is to inform the source node to migrate the specified container to its target node. It runs a Kafka Producer which publishes the message under the topic \texttt{$L_x$} where \texttt{$x$} stands for the host node ID.

\item \textit{Optimizer:} The primary container placement problem is modelled as an optimization problem with runtime metrics of the container as its input. Genetic algorithm is selected as an optimization algorithm to find the most stable and optimal container to node placement. 
It is based on the evolutionary computation, intimidating the process of biological reproduction and natural selection of the fittest individual. The algorithm starts by creating a huge population of randomly generated chromosomes. Each chromosome is an array of values denoting the \texttt{Node ID}. Consider $k$ containers running in the cluster represented by a tuple $(C_1,C_2,C_3....C_k)$.  For \texttt{N} worker nodes, the chromosome $C$ is encoded as \texttt{k} element array, \\
\begin{equation}
\label{eq}
C: [ M_{i} , M_{j} , M_k  ... M_{l}] \hspace{1.2em}
\forall {(i,j,k,l)} \in N
\end{equation}
where the index $i$ of the chromosome represents the container $C_i$ and the value at the index $i$ represents the random node ID from a set of $N$ machines. In order to assess the population, a $fitness function$ is defined, which quantifies the fitness of an individual chromosome. The algorithm begins by choosing a set of the randomly generated chromosome, that serves as the initial population (Generation 1). The process is followed by a  series of operations on chromosomes such as cross over and mutation. The algorithm computes the fitness of chromosomes in each generation and fittest \textit{elite} chromosomes are passed over to the next generation. 
\\
We specify $\mu_{rn}$ as the $r^{th}$ resource utilization of a node $n$. The number of containers running on node $n$ is denoted by $C_n$. We define $m\mu_{rn}$ as mean resource utilization of $r^{th}$ resource on node $n$ using equation \ref{eq1} :
\begin{equation}
\label{eq1}
    m\mu_{rn}=\frac{\sum_{C_n} \mu_{rn}}{C_n}
\end{equation}
We define Stability metric $S$ as variance of the mean utilization of resource $m\mu_{r}$ across all nodes in a cluster by equation \ref{eq2}.
\begin{equation}
\label{eq2}
S = \sum_{r} \sum_{n} \big ( \big (m\mu_{rn} - \sum m\mu_{rn} \big )^2 \big )
\end{equation} 
For every chromosome, we compute the metric $S$, which represents the overall variance in the resource usage across different nodes. The objective of our \textit{Optimization} function is to find an optimal container to node placement. However, the optimal placement schedule may result in a huge number of container migrations. For every chromosome, we compute the migration count $d^{MIG}$ as the distance between the \textit{initial} placement and the chromosome. We define \textit{initial} placement as the string where each index $i$ represents the host node id of container $C_i$. Equation \ref{eq3} computes the distance $d ^ {MIG}(x,y)$ between two strings $x$ and $y$, each of length $n$, by counting the number of mismatches among their pair of values.
\begin{equation}
\label{eq3}
d ^ {MIG}(x,y) = \sum_{k=1}^{n} \big [ y_{i,k} \neq y_{j,k} \big ]
\end{equation} 
We compute $S$ and $d_{MIG}$ for every chromosome across the entire population. To make the values comparable across the population, we normalize these values. $S_n$ and $d^{MIG}_n$ represents the normalized values for each of the chromosomes in the population. Finally in equation \ref{eq4}  we define the Fitness function $f$ used in Genetic algorithm as the weighted combination of $S_n$ and $d^{MIG}_n$.
\begin{equation}
\label{eq4}
\scalebox{1.2}{$ f = \alpha \times S_n + ( 1 - \alpha) \times d^{MIG}_n$}
\end{equation} 

where $\alpha$ is the tunable parameter, which indicates the trade-off between variance and migrations. If $\alpha$ is 0, the optimization function would only attempt to reduce variance, causing higher migrations. If $\alpha$ is 1, the optimization function would only reduce migrations at the cost of higher variance. Therefore we need to find the optimal value of $\alpha$, to achieve reduced variance with minimum migrations. Figure \ref{fig:alpha} shows the trade-off between Stability metric $S$ and migrations $d^{MIG}$ by varying the value of $\alpha$. The experiment was conducted by executing various types of containers in the Swarm cluster. The migration values are normalized for easy interpretation. We note that by keeping $\alpha = 0.85$, we can reduce variance as well as the number of migrations. The rest of the experiments have been conducted using $\alpha$ as 0.85. However, $\alpha$ can be tuned based on the desired scheduling objective. 
\end{enumerate}
\item \textit{Worker Nodes}: The containers run on the worker nodes. The components of the worker node are shown in Figure \ref{fig:worker}. The worker node has three major components:  \setlength{\parskip}{\baselineskip}
\begin{enumerate}
	\setlength \itemsep{1em}
	\item \textit{Stats Producer}: Its primary role is to fetch the metadata and runtime metrics of the container. The containers are profiled using a wide range of runtime metrics. These include CPU, memory, block I/O, network, etc. The metadata includes the container ID, node ID, resources allocated to the container, etc. The Stat Producer internally executes a Kafka Producer which publishes messages under the topic \texttt{$M_x$}, where \texttt{$x$} stands for worker node Id. These messages are consumed at the \textit{Manager node} by \textit{Stats Consumer}.   
	\item \textit{Result Consumer}: It is responsible for fetching the migration information from the \textit{Manager} and passing it to the \textit{Migration Module}. It internally executes a Kafka Consumer which subscribes to the topic \texttt{$L_x$} and enables it consume messages produced by the \textit{Result Producer} of the Manager node.
%	The migration information is forwarded to  Migration Module.    
	\item\textit{ Migration Module}: It is the principal component of each worker node, which is primarily responsible for issuing the migration instruction and carrying out the checkpoint-restore operation for the container. It receives information consisting of container ID and the target node. Upon receiving the information, it carries out the steps to perform the container migration to the target node. The steps to perform container migration along with two approaches are discussed in section \ref{sect:migration}. Container migration is a complex task involving significant use of network bandwidth. Figure \ref{fig:timedis}  shows the distribution of time taken for various steps involved in container migration. It can be deduced from the figure that significant time is required for \texttt{Docker commit} operation. We discuss the factors affecting the container migration in section \ref{sect:factorschk}.
\end{enumerate}
\item \textit{Private Container Registry}: We deploy a private Docker registry to speed up the process of container migration as discussed in  Approach 2.
\end{enumerate}
\begin{figure}
\hspace{5mm}
	\includegraphics[scale=0.25]{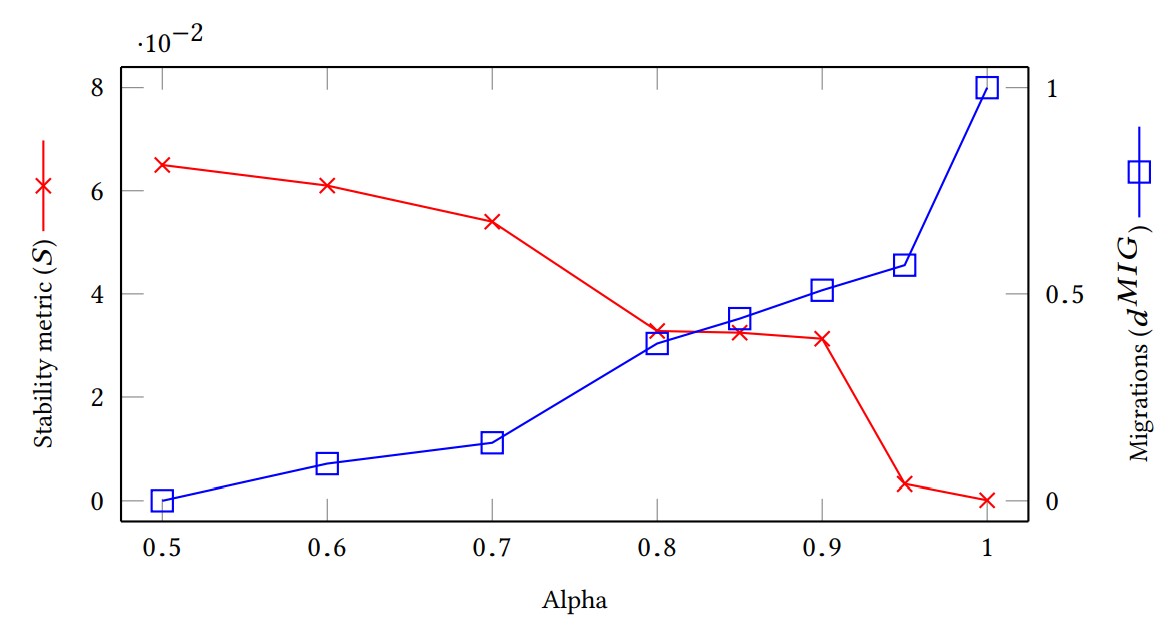}
	\captionsetup{font=scriptsize} \caption{Trade-off between Stability $S$ and Migrations $d^{MIG}$ by varying $\alpha$}
	\label{fig:alpha}
\end{figure} 
\begin{figure}
\hspace{12mm}
	\includegraphics[scale=0.25]{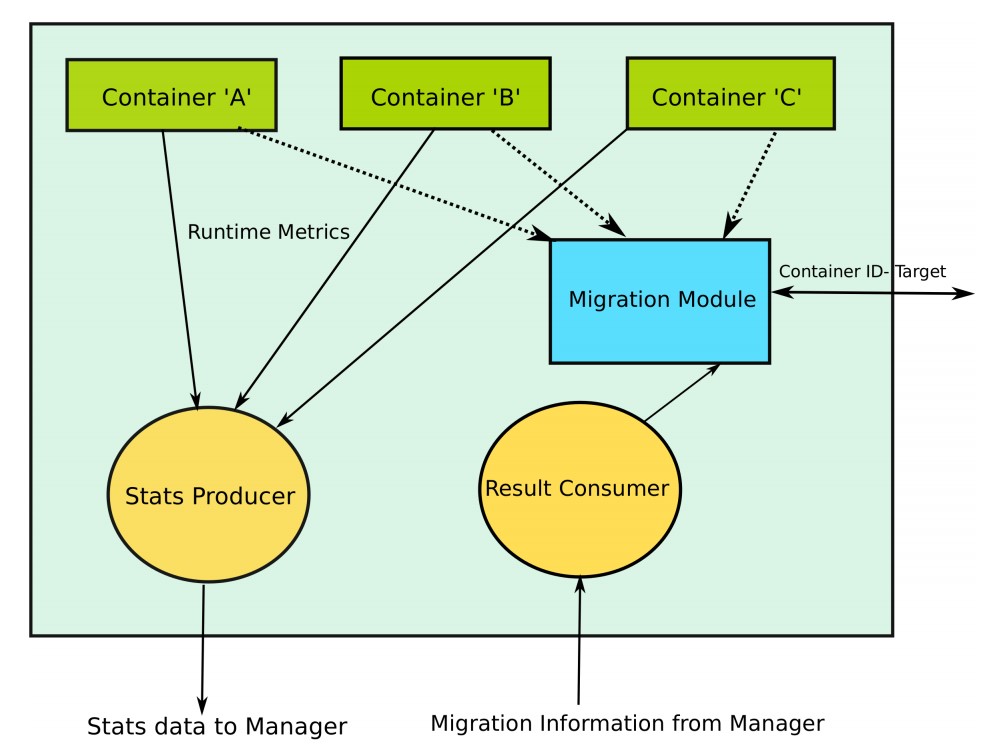}
	\captionsetup{font=scriptsize,skip=1pt} \caption[]{Worker Node}
	\label{fig:worker}
\end{figure}
\subsection{Tunable parameters in our framework}
\label{sect:tunable}
Our C-Balancer framework requires a few parameters to be tuned based on the scheduling objective. 
\begin{itemize}
	\item \textit{Alpha} ($\alpha$): The parameter decides the trade off between stability $S$ and migrations $d ^ {MIG}$ in the fitness function $f$. $\alpha$ can be tuned based on the scheduling objective that can range from minimized migrations to maximized stability.   
	\item\textit{Genetic algorithm}: Tunable parameters includes cross over probability, mutation probability, elitism count, size of population  and number of iterations. These parameters can be  adjusted after evaluating the performance of the algorithm in a few trials.
	\item \textit{Container performance metrics}: The parameters are used to profile of the container processes. They include parameters which profile CPU, memory, network, block I/O, etc for a container. We collect these performance parameters using cgroup. Another tunable parameter associated with performance metrics is container profiling rate. We observed that profiling performance parameters involve a very low overhead. In our experimentation, the containers are profiled every 5 seconds.       
	\item \textit{Frequency of invocation of $Optimizer$}: The frequency indicates how often $Optimizer$ is called to re-balance the containers across the worker nodes. We observed that a careful selection of parameter is necessary to avoid frequent migrations and instability. The frequency of invocation of the $Optimizer$ must be  at least inversely proportional to the time taken to migrate a container. A frequency higher than this would result in higher system instability and increased number of s migrations.
\end{itemize}

\section{Evaluation}
\label{sect:eval}
\subsection{Comparison of the two approaches for file system synchronization}
\label{sect:comptwo}

\begin{figure}
	\begin{tikzpicture}
	\begin{axis}[
	ybar stacked,
	width=6cm,
	height=4.3cm,
	legend style={
		at={(-0.9,0.5)},
		anchor=west,
		legend columns=1,
		font=\tiny,
		mark options={scale=1},
		/tikz/every even column/.append style={column sep=0.1cm}
	},
	xtick=data,
	symbolic x coords={
		Tomcat,
		Redis,
		Pi,
		Memcached,
		Owncloud,
		Httpd,
		Stream},
	ymin=0,ymax=20,
	ytick={0,2,4,...,20},
	area legend,	
	enlarge x limits={0.1},
	tick label style={font=\tiny},
	ylabel style ={align=center, font=\tiny},
	ylabel={Time (in seconds)},	
	bar width=12pt,
	]
	\addplot [hous,fill,postaction={
		pattern=north west lines
	}] coordinates {
		(Tomcat,1.18)
		(Redis,1.15)
		(Pi,1.106)
		(Memcached,1.14)
		(Owncloud,1.035)
		(Httpd,1.199)
		(Stream,1.18)
	};
	\addplot [farming,fill,postaction={
		pattern=north east lines
	}] coordinates {
		(Tomcat,4.04)
		(Redis,0.06)
		(Pi,0.416)
		(Memcached,0.017)
		(Owncloud,1.694)
		(Httpd,0.205)
		(Stream,0.872)
	};
	\addplot [green,fill,postaction={
		pattern=north west lines
	}]  coordinates {
		(Tomcat,3.83)
		(Redis,6.16)
		(Pi,6.512)
		(Memcached,5.59)
		(Owncloud,6.06)
		(Httpd,5.96)
		(Stream,5.36)
	};
	\addplot[trans,fill,postaction={
		pattern=north east lines
	}] coordinates {
		(Tomcat,3.305)
		(Redis,2.68)
		(Pi,2.6)
		(Memcached,2.2)
		(Owncloud,4.7)
		(Httpd,2.4)
		(Stream,1.17)
	};
	\addplot[aku,fill,postaction={
		pattern=north west lines
	}] coordinates {
		(Tomcat,0.253)
		(Redis,0.14)
		(Pi,0.207)
		(Memcached,0.144)
		(Owncloud,0.278)
		(Httpd,0.157)
		(Stream,0.206)
	};
	\addplot [other,fill,postaction={
		pattern=north east lines
	}] coordinates {
		(Tomcat,1.07)
		(Redis,0.22)
		(Pi,0.093)
		(Memcached,0.013)
		(Owncloud,0.597)
		(Httpd,0.06)
		(Stream,0.213)
	};
	\addplot [indu,fill,postaction={
		pattern=north west lines
	}]  coordinates {
		(Tomcat,0.132)
		(Redis,0.2)
		(Pi,0.537)
		(Memcached,0.23)
		(Owncloud,0.489)
		(Httpd,0.46)
		(Stream,0.244)};
	\addplot [water,fill,postaction={
		pattern=north east lines
	}]  coordinates {
		(Tomcat,2.6)
		(Redis,0.129)
		(Pi,0.28)
		(Memcached,0.296)
		(Owncloud,0.33)
		(Httpd,0.242)
		(Stream,0.25)
	};
	\addplot [techinfra,fill,postaction={
		pattern=north west lines
	}] coordinates {
		(Tomcat,2.104)
		(Redis,0.839)
		(Pi,1.314)
		(Memcached,1.004)
		(Owncloud,2.04)
		(Httpd,1.009)
		(Stream,1.03)
	};
	\legend{
		Create Checkpoint,
		Compress Checkpoint,
		Commit File System,
		Push Image,
		Transfer Checkpoint,
		Extract Checkpoint,
		Pull Image (Layers Present),
		Create Container,
		Restore Container	
	}
	\end{axis}
	\end{tikzpicture}
	\captionsetup{font=scriptsize,skip=1pt}
	\caption[]{Time taken to perform each step of container migration for popular Docker images.}
	\label{fig:timedis}
	\vspace{-3mm}
\end{figure}
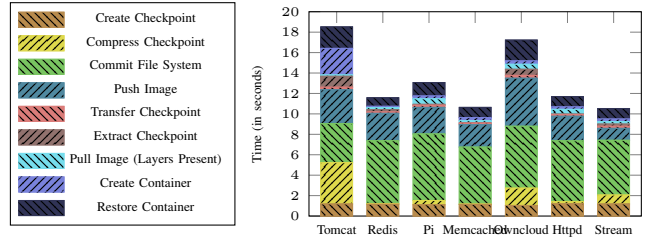

\begin{figure}
	\includegraphics[scale=0.2]{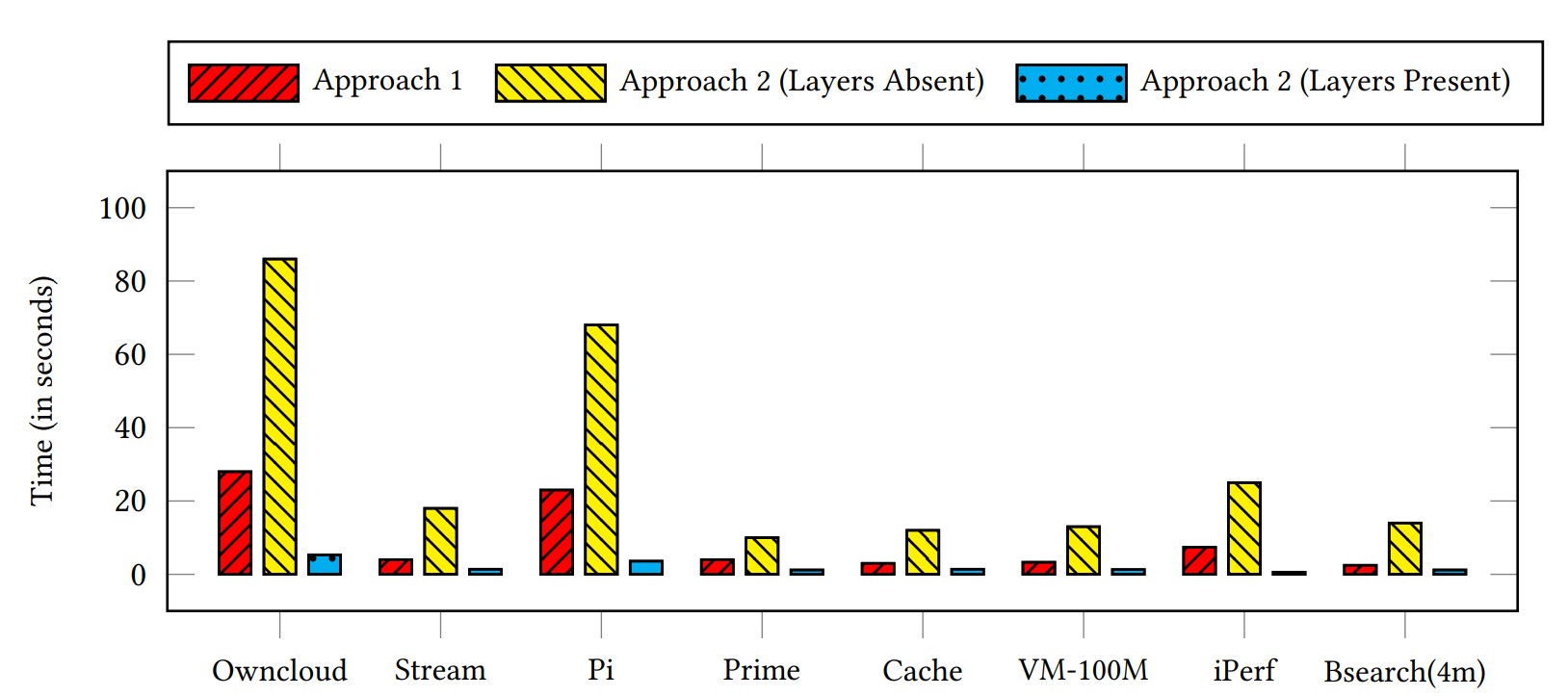}
		\captionsetup{font=scriptsize} \caption[]{Comparison of the two approaches used for synchronizing the file system of container. Approach 1 exports the file system of container as a compressed file. Approach 2 (Layers Absent) shows the time taken to commit file system when no image layer is present in private registry. Approach 2 (Layers Present) shows time taken when the container image layers (except the \texttt{init} layer) are present the in registry.}
	\label{fig:pushpull}
	\vspace{-4.8mm}
\end{figure}

Figure \ref{fig:timedis} shows the time required for every step of container migration. The stacked bar graph shows that among all steps, the \texttt{commit} operation takes higher time to complete. The figure also reveals that steps such as creating/restoring the checkpoint take substantial time. Moreover, the time required for file system synchronization (\texttt{push/pull}) is significantly less and is attributed to the benefits of using Approach 2.
Figure \ref{fig:pushpull} compares the two approaches of file system synchronization for popular Docker images. We also evaluate the approach by containerizing the benchmark programs of Stress-NG and iPerf benchmark. It can be inferred from the graph that the Approach 1 takes a significantly higher time. The Approach 2 (layers absent) indicates that the image layers are not available in the private registry. As a result, all the image layers are transferred to the registry and pulled back at the target node. Even though both the approaches involve the transfer of the entire image, the latter approach takes far more time than the former. The former approach transfers the image layers once (host to target), whereas the latter approach transfers the image layer twice, first from the host to the registry, and second from the registry to the target.
The Approach 2 (Layers Present) is the fastest among all the methods, with the consideration that the read-only image layers are already present in the private registry. This approach is based on the assumption that with the due passage of time, most of the read-only layers of Docker images gets added up to the registry. The approach may seem optimal and unreal, but as the containers continue to migrate, eventually the registry gets filled up, speeding up the process of container file system synchronization.

\subsection{Factors affecting checkpoint time}
\label{sect:factorschk}
\begin{figure}
	\includegraphics[scale=0.23]{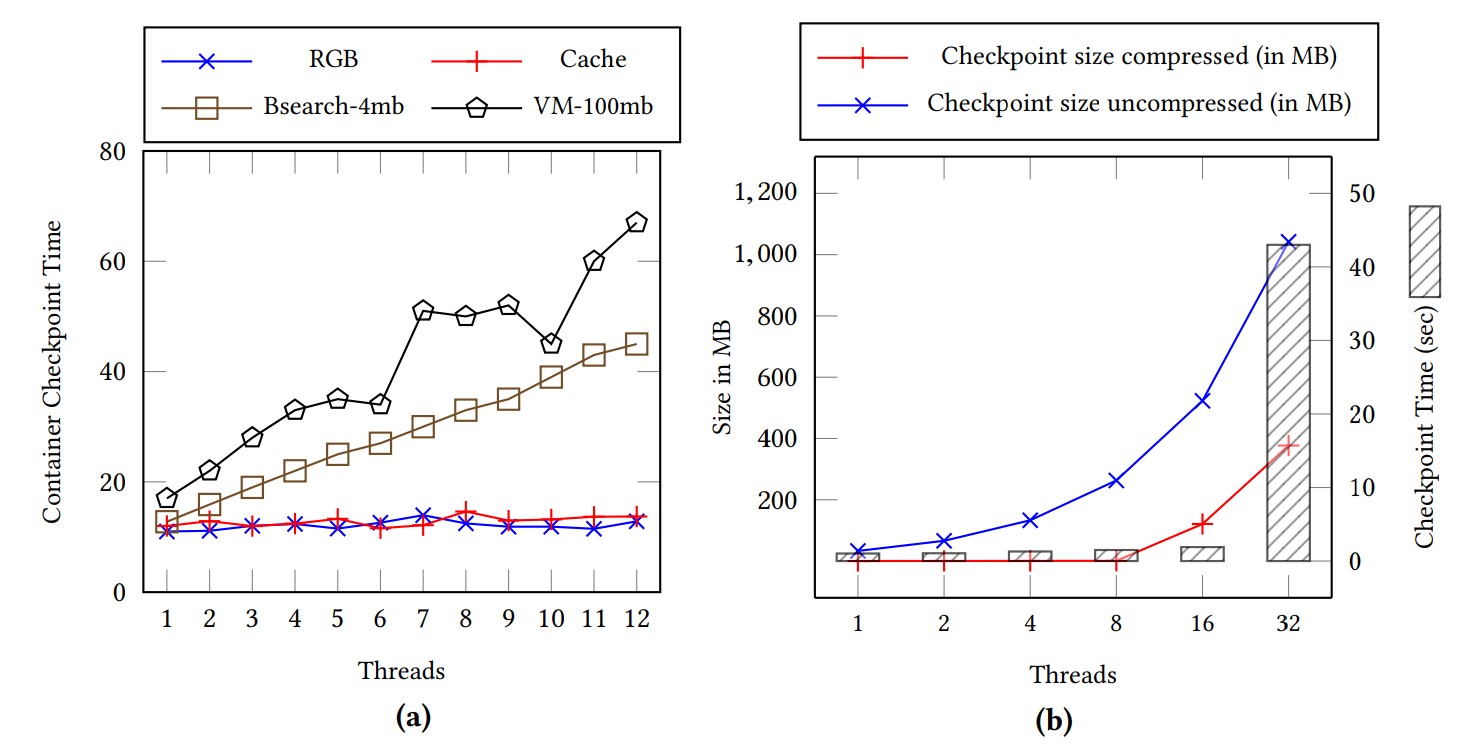}
		\captionsetup{font=scriptsize} \caption{ Figure (a) shows the effect of the number of threads on the checkpoint time of various containers. Figure (b) shows the variation of the checkpoint time and checkpoint data size with increase in number of threads for \texttt{Qsort} program (Stress-NG Benchmark) }
	\label{fig:combined}
	\vspace{-3mm}
\end{figure}

In this subsection, we discuss the factors impacting the container checkpoint time. We observed that the checkpoint time is mainly affected by the total memory footprint of the processes running inside the containers. Memory check-pointing involves dumping the pages owned by the process and its associated metadata(page table) as a binary file. We run \texttt{Qsort} program of Stress-NG \cite{Stress-NG} benchmark as a container to measure the checkpoint time and checkpoint data size. Figure \ref{fig:combined}(b) shows that the checkpoint size (uncompressed) increases exponentially with the number of threads. This rise is because, an increase in the number of threads  results in increased memory footprint of the processes. This increased memory footprint is reflected in the increased checkpoint size and checkpoint time. Figure \ref{fig:combined}(b)  also highlights the importance to compress the checkpoint before transferring to the target node. The checkpoint size (compressed) is far lesser than checkpoint size (uncompressed), and therefore substantially improve the transfer  time and network bandwidth utilization.

To study the impact of the number of threads on the checkpoint time, we experimented it using four distinct programs of Stress-ng benchmark: \textit{RBG} (a CPU intensive process with very low memory footprint), \texttt{Cache} (process that stresses cache with very low memory footprint), \textit{Bsearch - 4m} (program to perform binary search on a sorted array of size 4m, having slightly higher memory footprint and CPU intensive) and \textit{VM- 100m} (a highly memory intensive program with 100 Mbytes memory footprint per thread). We run each of these programs in a container and vary the number of threads. From figure \ref{fig:combined}(a), we observe that the checkpoint time is not affected for \textit{RBG, cache} containers. However, it increases linearly for \textit{Bsearch} and drastically for \textit{VM} containers. The sharp increase  for \textit{VM-100mb} is due to an increase in the memory footprint of the container processes.

\subsection{Evaluation using Stress-NG benchmark}
\begin{figure*}
\includegraphics[scale=0.4]{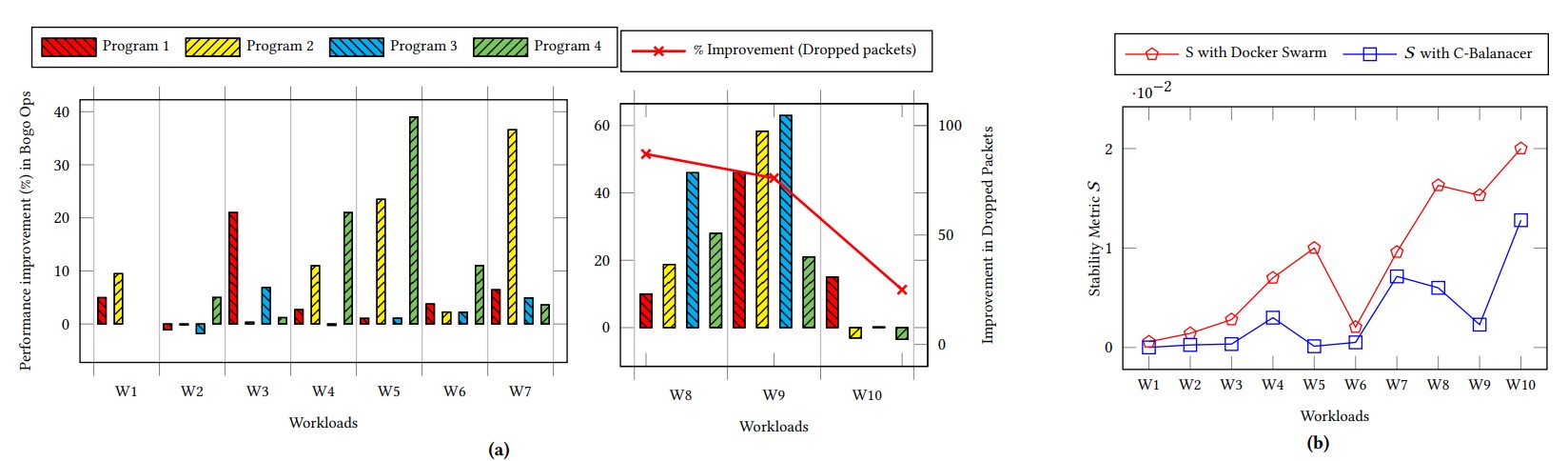}
	\captionsetup{font=scriptsize} \caption[]{Figure (a) shows the performance improvement in terms of Bogo Ops with default Docker Swarm scheduler. Figure (b) shows the comparison of Stability metric $S$ for Docker Swarm and our framework \textit{C-Balancer}.}
\label{fig:mainres}
\vspace{-2mm}
\end{figure*}
\begin{table}
	\centering
	\scriptsize
	\begin{tabular}{|c||c|c|}
		\hline
		\textbf{Category} & \textbf{Description}\\
		\hline\hline
		Manager node & 1\\
		Worker nodes & 14 \\
		Kafka Broker Nodes & 2 \\
		Zookeeper Node & 2 \\
		Private Registry & 1 \\  
		Linux Version &  Linux BOSS 3.16.0-4-amd64 \cite{bossmool} \\
		Docker  & Experimental Version: 1.13.0 \\
		Kafka & Version: 2.11 \\
		CRIU & Version: 3.4 \\ 
		Nodes Configuration & 4 Cores and 4GB memory\\ 
		\hline
	\end{tabular}
\captionsetup{font=scriptsize,skip=1pt}
\vspace{1mm}	
\caption{System Configuration}
\label{tab:freq}
\vspace{-4mm}
\end{table}
In this subsection, we evaluate our prototype and show how \textit{C-Balancer} can be configured to achieve performance improvement. We evaluate our proposed approach using Stress-NG \cite{Stress-NG} and iPerf \cite{iperf3} benchmark. Stress-NG benchmark is a tool that stresses and measures various subsystems such as CPU, memory, cache, disks etc. iPerf is a tool, used to measure the end to end network bandwidth. The iPerf client can be configured with varying request rate to stress the iPerf server and network components of the system. To stress the system resources, we chose the following programs of Stress-NG benchmark.

\begin{itemize}
\item \textit{CPU}: The program  stresses the  CPU and its various functional units. Example: \textit{pi, RGB, matrixprod}, \textit{crypt, prime, stats,} etc. These programs exercise the CPU extensively. 
\item \textit{Cache}: The program performs widespread and random memory reads and writes to thrash the CPU cache.
\item \textit{Memory Stress}:  The program stresses the various subsystems of memory. Example: \textit{vm} calls mmap or munmap \cite{mmap} function aggressively to exercise the memory resources of system.
\item \textit{General Programs}: These programs measure the performance of application by stressing most of the subsystems. Example: \textit{Bsearch}, a binary search program on a sorted array, \textit{Tsearch}, a program to perform insert/delete/search on a binary tree, \textit{Qsort}, a program that sorts the integers using quick sort. 
\end{itemize}
  
The benchmark programs are run as Docker container in the Swarm cluster. Each container is run for a total of 120 seconds. We use \texttt{--timeout} to control the process execution time. The benchmark shows the throughput results using \texttt{bogus operation per second} (Bogo Ops). It provides the rough notion of performance and is not comparable across different stressors programs. We use \texttt{--metrics-brief} to fetch the performance results for Stress-NG benchmark.

Our setup consists of nodes (virtual machines) that runs on BOSS MOOL(Minimalistic Object Oriented Linux) \cite{bossmool} operating system. However, our implementation is generic enough to run on other Linux distributions. Table \ref{tab:freq} shows the detailed system configuration. 
\begin{table}
	\scriptsize
	\begin{tabular}{|c||c|c|c|c|}
		\hline
		\textbf{Workload} & \textbf{Program 1} & \textbf{Program 2} & \textbf{Program 3} &\textbf{Program 4} \\
		\hline\hline
		W1 & RGB & Bsearch (4M) & RGB & Bsearch (4M) \\
		W2 & Prime & Bsearch (4M) & RGB & Cache \\
		W3 & Cache & Pi & Cache & Prime \\
		W4 &Prime & Stream & Queens & Cache \\
		W5 &psi & Stream & Prime & Stream \\
		W6 & Prime & Bsearch (4M) & Crypt & Cache \\
		W7 & Crypt & Tsearch (4M)& Queens & Cache \\
		W8 & iPerf (100 M) & Stream & iPerf (150 M) & Cache \\
		W9 &iPerf (100 M) & VM (50 MB) & iPerf (150 M) & Stream \\
		W10 & iPerf (100 M) & VM (50 MB) & Queens & Cache\\
		\hline
	\end{tabular}
\centering
\captionsetup{font=scriptsize,skip=1pt}
\vspace{1mm}	
\caption{Workload mix of benchmark programs from Stress-NG and iPerf. The mix is chosen to stress various resources of the system including CPU, cache, memory, network and I/O. The  \textit{M} denotes speed in Mbps}
%\vspace{-3mm}
\label{tab:wrokload}
\vspace{-5mm}
\end{table}

For evaluation, we create a workload mix of processes from Stress-NG and iPerf benchmarks, as shown in Table \ref{tab:wrokload}. The programs are chosen so as to cover and stress all the resources in the system. The benchmark programs are run with \texttt{replication-factor 7}. The containers are launched in the same order as shown in the table. For instance, in \texttt{W1}, seven replicas of \textit{RGB} containers are launched followed by seven replicas of \textit{Bsearch (4m)} and so on. We deliberately create certain workload mix and decide the container launch order, to cause maximum resource contention. For instance, in \textit{W2}, there is an equal number of running containers after the launch of Bsearch and \textit{Prime} containers respectively in that order. When \textit{RGB} and \textit{Cache} containers are launched next, the containers are placed randomly by Swarm. So a \textit{Cache} container may be co-running container with either \textit{Bsearch} or \textit{Prime} container. Thus there is a certain randomness in the scheduling decision by Swarm, which can cause significant resource contention.  We use \texttt{(--restart-condition none)} for launching containers as services in the cluster. The tweak is done to prevent Swarm from re-starting the containers, as the containers are stopped during the checkpoint operation. We compare our proposed approach \textit{C-Balancer} with default Docker Swarm scheduling.

Figure \ref{fig:mainres}(a) shows the performance improvement for each of the workload mixes. The performance improvements are averaged for five runs in each of the experiment. Figure \ref{fig:mainres}(b) analyses the Stability metric $S$ for workload mixes using Docker Swarm and our proposed \textit{C-Balancer} framework. The Stability metric $S$ represents the variance of the resource utilization in the cluster. We observe the maximum average performance improvement of 58 \% for workload $W9$. The improvement is attributed to a significant decrease of around 85 \% in the value of $S$. The workloads which have a lesser reduction in $S$ value, have a lesser performance improvement.

Also, we observe an overall reduction of 61 \% on average in the value of $S$ across all workloads. The decrease in $S$ is paramount as it substantially reduces the resource contention, thereby stabilizing the cluster. We also observed that the  memory, cache and network intensive containers show a tremendous performance improvement as against the CPU intensive containers. This is evident, as the resource contention caused due to memory, cache and network exceed far more than that of a CPU. Additionally, for network containers, we also evaluated the change in the percentage of dropped packets with \textit{C-Balancer}. With C-Balancer, we are able to achieve a maximum reduction of about 58 \%  in the number of dropped packets as against Docker Swarm. The experiments are performed with independent loosely coupled containers, scheduled on the Swarm Cluster. However, we do not analyse and speculate the performance of C-Balancer in Kubernetes \cite{google}, when multiple tightly coupled containers are encapsulated in a pod and launched as a service.

\section{Related Work}
\label{sect:related}
Virtualization serves as one of the major backbones for cloud computing systems and as a
result, VMs have been studied extensively for a long time. VM provides benefits such as hardware independence, isolation and security. However, they have significantly higher performance overhead, longer startup times and limited scalability. Recently container based OS virtualization \cite{soltesz2007container} has gained significant attention. There has been significant work done regarding performance analysis of virtual machine over containers \cite {soltesz2007container}, \cite{xavier2013performance} \cite{ruiz2015performance}, \cite{felter2015updated}.
Several issues concerning the containers ranging from provisioning time \cite{hegde2016scope} to high availability \cite{nathan2017comicon}, \cite{li2015leveraging} are addressed. \citeauthor{nathan2017comicon} presents the system for managing the Docker images in a distributed manner. \citeauthor{nadgowda2017comparing} \cite{nadgowda2017comparing} describes the various scaling methods for containers. The paper describes the cold scale, hot scale and warm scale methods which effectively improves the container provisioning time. \citeauthor{hegde2016scope} builds the model to predict the container provisioning time using load (request per second) as an input. However, it does not consider other container runtime-metrics and their proposed model can be used for some select applications. \citeauthor{li2015leveraging} \cite{li2015leveraging} focuses on providing high availability for a cloud application, by continuously checkpointing the container at regular intervals. However, they do not focus on file system synchronization for container migration. The approach simply provides backup to the container in the event of a failure.
\cite{kaewkasi2017improvement} and \cite{mao2017draps} are the two early works that address the concerns of container scheduling. \citeauthor{kaewkasi2017improvement} \cite{kaewkasi2017improvement} used Ant Colony Optimization technique for container scheduling. However, they do not evaluate the approach extensively and highlight the data collection methods and overheads involved. \citeauthor{mao2017draps} \cite{mao2017draps} addresses the problem of resource awareness for container placement. They identify the dominant resource of a container for deciding the schedule. However, their scheduling decision does not consider other co-running containers in the cluster. \citeauthor{ma2017efficient} \cite{ma2017efficient} is one of the first work which presents efficient service hand-off for Docker containers. However, they do not address the concerns of container scheduling algorithms.
\label{sect:concl}
In this paper, we propose C-Balancer, a novel framework for profiling and scheduling the  containers in the cluster environment. The key idea behind the approach is to re-balance the containers across multiple nodes using container runtime metrics. The paper describes the container migration in detail. It also proposes two approaches for the container file system synchronization. From our prototype evaluation, we observed  that C-Balancer reduces the variance in the resource utilization across the cluster by 60\% on average. The improved resource utilization translates to performance improvement of upto 58\% for a workload mix of Stress-NG and iPerf benchmark programs. Our framework can be easily adapted for different scheduling criteria and can be tuned for different workloads environment. Our proposed framework is easily scalable to a large number of nodes, and the optimizer can leverage the power of Graphical Processing Units (GPU) for making faster scheduling decisions.
The current design of our framework uses very few parameters. In future, we plan to incorporate hardware performance counters for better scheduling decisions. Further, the containers are prone to various kinds of attack in the public cloud. Early prediction of an attack using the MOOL \cite{bossmool} message filters and timely container migration using our framework is an exciting research direction, which we leave as future work.
\printbibliography
\end{document}